\documentclass[prl,showpacs,twocolumn,aps]{revtex4}%
\usepackage{color}
\usepackage{graphicx}
\usepackage{epsfig}
\usepackage[english]{babel}
\usepackage{amsmath}
\usepackage{amsfonts}
\usepackage{amssymb}%
\newcommand{\be}{\begin{equation}}
\newcommand{\ee}{\end{equation}}
\newcommand{\bel}[1]{\begin{equation}\label{#1}}
\newcommand{\bea}{\begin{eqnarray}}
\newcommand{\eea}{\end{eqnarray}}
\newcommand{\balign}{\begin{align}}
\newcommand{\ealign}{\end{align}}
\newcommand{\ba}{\begin{array}}
\newcommand{\ea}{\end{array}}
\newcommand{\bfig}{\begin{figure}}
\newcommand{\efig}{\end{figure}}

\newcommand{\eref}[1]{(\ref{#1})}

\allowdisplaybreaks

\newcommand{\exval}[1]{\mbox{$\langle \, {#1}\, \rangle$}}

\newcommand{\rme}{\mathrm{e}}

\newcommand{\half}{\frac{1}{2}}

\newcommand{\comm}[2]{\mbox{$[\,{#1}\,,\,{#2}\,]$}}

\begin{document}

\title{Conformal invariance in driven diffusive systems at high currents}
\author{D. Karevski$^1$, G.M. Sch\"utz$^{1,2}$}
\affiliation{(1) Institut Jean Lamour, dpt. P2M, Groupe de Physique Statistique, 
Universit\'e de Lorraine, CNRS UMR 7198, B.P. 70239, F-54506 Vandoeuvre les Nancy Cedex, France}
%\author{G.M.~Sch\"utz}
\affiliation{(2) Institute of Complex Systems II, Theoretical Soft Matter and Biophysics,
Forschungszentrum J\"ulich, 52425 J\"ulich, Germany}
%\affiliation{Interdisziplin\"ares Zentrum f\"ur Komplexe Systeme, Universit\"at
%Bonn, Br\"uhler Str. 7, 53119 Bonn, Germany}

\begin{abstract}
We consider space-time correlations in driven diffusive systems which undergo a 
fluctuation into a regime with an atypically large current or dynamical activity. 
For a single conserved mass we show that the spatio-temporal density correlations 
in one space dimension are given by conformally invariant field theories with 
central charge $c=1$, corresponding to a ballistic universality class with 
dynamical exponent $z=1$. We derive a phase diagram for atypical 
behaviour that besides the conformally invariant regime exhibits a regime of 
phase separation for atypically low current or activity. On the phase transition 
line, corresponding to typical behaviour, the dynamics belongs to the 
Kardar-Parisi-Zhang universality class with dynamical exponent $z=3/2$, 
except for a diffusive point with $z=2$. We demonstrate the validity
of the theory for the one-dimensional asymmetric simple exclusion process 
with both periodic and open boundaries by exact results 
for the dynamical structure in the limit of maximal current.   
\end{abstract}
%\date{\today }


\maketitle

An intriguing question is whether a many-body system far from thermal equilibrium 
that undergoes some big atypical fluctuation can be understood in terms of an 
upscaled description of ``normal'' spatio-temporal behaviour or whether it 
behaves in a qualitatively different fashion during this atypical fluctuation. 
The statistical properties of such fluctuations are also the fundamental
object of interest in large deviation theory which provides -- in equilibrium --
the foundations of thermodynamics in terms of thermodynamic ensembles.
Of course, one does not expect a unique answer to a question posed so 
generally, but some insight into large fluctuations 
without external trigger may be gained from considering
a reasonably wide, but still well-defined class of model systems, viz. 
stochastic lattice 
gas models that have served as paradigmatic models for non-equilibrium 
phenomena in the last decades \cite{Spoh91,Schu01,Scha10}.

These models have a stochastic particle hopping dynamics 
that mimics noise, interactions between particles frequently include a 
hard-core repulsion, and a bulk driving field or boundary gradients maintain a 
fluctuating non-equilibrium steady state with a non-zero locally 
conserved mass-current. The paradigmatic example is the asymmetric simple 
exclusion process (ASEP) in one dimension \cite{Spoh91,Schu01,Scha10,Derr07} 
where a lattice site can be 
occupied by at most one particle and particles jump randomly to their nearest 
neighbour sites, provided the target site is empty. A bias with jump 
rates $p$ to the right and $q$ to the left and/or open boundaries where 
particle exchanges with reservoirs take place create a non-equilibrium 
situation.

The question is whether in a driven lattice gas the spatio-temporal 
fluctuation patterns remain essentially unchanged during an untypical large 
fluctuation of the current (or more generally of the undirected jump 
activity of the particles), or whether correlations during such a fluctuation are 
qualitatively different from typical behaviour, which in one space dimension 
has recently been shown to be universal and within the scope of
the theory of non-linear fluctuating hydrodynamics \cite{Spoh14}
which predicts for one conserved current either diffusive behaviour with 
dynamical critical exponent $z=2$ or fluctuations in the
universality class of the Kardar-Parisi-Zhang (KPZ) equation with $z=3/2$.

The answer that we shall give is, in a nutshell, the following: 
If a system with hard-core repulsion that is typically in a
stationary state of spatially homogeneous density undergoes a large fluctuation 
into a regime of {\it low} current or jump activity then this is most likely 
realized not by fluctuations in the random time after which particles 
attempt to jump (upscaling), but by spontaneous phase 
separation into spatial domains of high and low density respectively.
Conversely, during a fluctuation into a regime of {\it high} current or 
activity the 
homogeneous stationary density is maintained, but a qualitative change of 
correlations between particle positions makes the atypical large fluctuation 
least unlikely and hence typically realizes it. More specifically, we assert
that these space-time correlations are universal and can be predicted for 
non-equilibrium particle systems in one space dimension from conformal field 
theory for two-dimensional equilibrium critical phenomena \cite{Card96,Henk99}.

The first answer concerning atypically low current or activity can be understood
by noting that in a region of high density fewer jumps will occur naturally due 
to the repulsive interaction, while in a low-density region fewer jumps occur
trivially because of the smaller number of particles, thus optimizing the 
probability for such untypical behaviour. This picture is well borne out by
the powerful machinery of Macroscopic Fluctuation Theory 
\cite{Bert15} which demonstrates for the ASEP with periodic
boundary conditions that phase separation sets in below 
some critical atypical current \cite{Bodi05,Espi13}. For open boundary conditions
a similar phenomenon occurs \cite{Bodi06,Beli13b,Laza15}. Such a 
dynamical phase transition was found also for an atypically low activity
in the symmetric simple exclusion process (SSEP) which has no hopping 
bias \cite{Leco12}.

On the contrary, for a fluctuation into a regime of 
high current or activity phase separation would be counter-productive
and one expects a homogeneous bulk density as in the typical steady state.
However, no theoretical framework for a quantitative description of this
regime, which is inaccessible to macroscopic fluctuation theory, 
has been given yet. It is aim of this work to establish that in 1+1
dimensions this regime can be described by conformal field theory (CFT). 

More precisely, we predict that in the stationary regime of an atypical 
speeding-up of particle hopping in an interacting system with a single locally 
conserved current the dynamical structure function 
$S(k,l,t) = \exval{(n_k(t)-\rho_k)(n_l(t)-\rho_l)}$, i.e., the 
stationary space-time correlations of the local particle numbers $n_k(t)$ 
with local average density $\rho_k$, has in a translation invariant setting 
with $x=k-l$ the universal scaling form
\be 
\label{1}
S(x,t) = \frac{C_1}{v_L^2 t^2} 
\frac{1-\xi^2}{(1+\xi^2)^2} + \frac{C_2}{(v_Lt)^{2\gamma}} \frac{\cos{[2(q^\ast x - \omega t)]}}{(1+\xi^2)^{2\gamma}}
\ee
with the scaling variable $\xi = (x - v_c t)/(v_L t)$ indicating dynamical
exponent $z=1$ rather than 2 or 3/2 for typical dynamics. The static
critical exponent $\gamma \geq 1$ depends on the
particle interactions, the collective velocity $v_c$, the Luttinger liquid 
time scale $v_L$, the wave vector $q^\ast$, and the oscillation frequency 
$\omega$ can be computed from the generally complex dispersion relation 
as discussed below, and $C_i$ are non-universal amplitudes that depend on 
the microscopic details of the model. Higher-order correlations are fully 
determined by a modified CFT with central charge $c=1$ of the Virasoro 
algebra, the difference to usual conformal invariance being the appearance 
of the collective velocity $v_c$ in the Galilei-shift of the space coordinate 
$x$ and the time-dependence of the oscillating part of the correlation 
function with frequency $\omega$.

\begin{figure}[ptb]
\begin{center}
\includegraphics[scale=0.36]{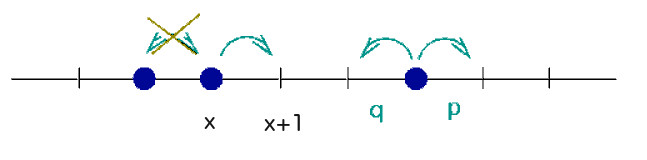}
\end{center}
\caption{(Color online) Schematic representation of the 
asymmetric simple exclusion process. 
A particle hops to the neighbouring site provided this target site is empty.
In the regime of maximal current the jumps to the left do not contribute to the 
statistical properties of the conditioned process.}%
\label{Fig:ASEP}%
\end{figure}

We prove this assertion for the ASEP (Fig. \eref{Fig:ASEP}) in the regime of 
maximal current, both for periodic and open boundary conditions, by using the 
exact mapping of the dynamics of the ASEP conditioned on an atypical current or 
hopping activity to a non-Hermitian 
%one-dimensional quantum problem, viz. the 
Heisenberg spin-1/2 
quantum chain in the ferromagnetic range with anisotropy parameter 
$\Delta \geq 0$ and imaginary Dzyaloshinskii--Moriya interaction \cite{Schu15a}
\bel{HSEPXXZ}
\tilde{H} = - \frac{w}{2}\sum_{k=1}^{L} \tilde{g}_k
\ee
with
\be 
\label{XXZhopping}
\tilde{g}_k =  2 v \sigma^+_{k} \sigma^-_{k+1} +
2v^{-1} \sigma^-_{k} \sigma^+_{k+1}+ \Delta ( \sigma^z_k \sigma^z_{k+1} - 1).
\ee 
Here 
\be 
\rme^{f} = \sqrt{\frac{p}{q}}, \, w=\sqrt{pq} \rme^{\mu}, \,  
v=\rme^{f+\lambda}, \, \Delta = \rme^{-\mu} \cosh{f} 
\ee 
and $\lambda$ is conjugate to the current $j$ and $\mu$ is conjugate to the activity 
$a$ on which we condition. The case $\lambda=\mu=0$ corresponds to the 
typical dynamics with stationary current $j= (p-q) \rho(1-\rho)$ and 
activity $a= (p+q) \rho(1-\rho)$ for particle density $\rho=N/L$. 
Positive $\lambda$ and $\mu$ correspond to high
current and activity resp. \cite{Chet14,Jack15}.

The ground state of the {\it Hermitian} spin-1/2 Heisenberg quantum chain 
is known to be described by CFT with central charge $c=1$ 
\cite{Baxt82,Card96,Henk99}.
In order to show that the non-Hermitian terms in \eref{XXZhopping}
lead to the modifications of CFT described above 
we focus on maximal current $\lambda\to\infty$. 
Following \cite{Popk10,Popk11}, the maximal current is 
realized by two mechanisms: The trivial speed-up of the jump frequency and 
a non-trivial building up of correlations. In order to extract the non-trivial 
part we rescale time by $p\rme^{\lambda+\mu}$ so that we are left with
\bel{H1a}
H  =  -  \sum_{k=1}^{L} \sigma^+_{k} \sigma^-_{k+1}
\ee
for periodic boundary conditions with $L$ sites.

The dynamical structure function in this limit was computed in \cite{Popk11} 
and represents a ballistic dynamical universality class with dynamical exponent
$z=1$ that was first studied by Spohn \cite{Spoh99}. The large-scale behaviour 
%of the dynamical structure function 
is given by the non-oscillating part in
\eref{1}. However, neither in \cite{Popk11} nor in \cite{Spoh99} the oscillating
contribution has been pointed out. In order to extract this oscillating 
part and uncover other hallmarks of conformal invariance we take the standard
approach by diagonalizing \eref{H1a} in terms of Jordan-Wigner fermionic
operators \cite{Lieb61}. Fourier transformation
with momentum $p$ then yields 
%in terms of Pauli matrices $\sigma^\pm = (\sigma^x \pm i \sigma^y)/2$ and 
%particle projector $\hat{n} = (1 - \sigma^z)/2$ by
\bel{H1}
H = 
\sum_{p=1}^L (\mathcal{P}^- \epsilon_p \hat{c}^\dagger_p \hat{c}_p + \mathcal{P}^+
\epsilon_{p-\half} \hat{c}^\dagger_{p-\half} \hat{c}_{p-\half})
\ee
in the graded Fock space build by the free fermion
creation operators $\hat{c}^\dagger_p$ and annihilation operators $\hat{c}_p$
in the even particle sector and $\hat{c}^\dagger_{p-1/2}$, $\hat{c}_{p-1/2}$ 
in the odd sector. The single-particle energy is
\bel{energy}
\epsilon_p := - \rme^{-\frac{2\pi i p}{L}} .
\ee

Remarkably the ground state for $0 \leq N \leq [L/2]$ particles, 
which is defined by having the lowest real part of the eigenvalues of \eref{H1},
is the same as for the Hermitian case with
ground state energy per site given by
\bel{eps0}
- e_0 = \frac{v_F}{L\sin{(\pi/L)}} = v_F\left(
\frac{1}{\pi} + \frac{\pi}{6L^2} + O(L^{-4}) \right)
\ee
where $v_F=\sin{(\pi\rho)}$ is the Fermi velocity.
The energy gaps, however, are in general complex. The 
lowest gap (with smallest real part) is given by
$\delta = \epsilon_{(N+1)/2} - \epsilon_{(N-1)/2} = 
2 \sin{(\pi/L)} (\sin{(\pi\rho)} + i \cos{(\pi\rho)})$
with real part
\bel{gapreal}
\Re(\delta) = v_F \frac{2\pi}{L} \left( 1 + O(L^{-2}) 
\right).
\ee
%The leading finite-size corrections are consistent with CFT with 
%central charge $c=1$ and smallest critical exponent $x=1$.

For the dynamical structure function at density $\rho$ 
the Wick theorem gives
\bea
S_\rho (n,m,t) & = & \langle c^\dagger_n(t)c_m(0)\rangle\langle c_n(t)c^\dagger_m(0)\rangle
\nonumber \\
\label{Wickopen}
& &  - 
\langle c^\dagger_n(t)c^\dagger_m(0)\rangle \langle c_n(t)c_m(0)\rangle 
\eea
where the second part vanishes in the periodic system due to particle 
number conservation. The
time-dependent operators are defined by
$X(t) := \rme^{Ht} X \rme^{-Ht}$
which leads to
%\bel{FTtime}
$\hat{c}_{p}(t) = \rme^{-\epsilon_p t} \hat{c}_{p}$ and
$\hat{c}_{p}^\dagger(t) = \rme^{\epsilon_p t} \hat{c}_{p}^\dagger$
and allows us to compute straightforwardly the basic correlators in \eref{Wickopen}.

In terms of the function 
\bel{fdef}  
f_{L,N}(r,t) := \left\{\ba{ll} 
\displaystyle
\frac{1}{L} \sum_{p=-(N-1)/2}^{(N-1)/2} \rme^{\frac{-2\pi i}{L} p r +\epsilon_p t} 
& N \mbox{ odd} \\[8mm]
\displaystyle
\frac{1}{L} \sum_{p=-N/2+1}^{N/2}  
\rme^{\frac{-2\pi i }{L}(p-\half) r+\epsilon_{p-\half} t} & N \mbox{ even}.
\ea \right.
\ee 
we find the exact result
\bea
\label{basiccorr1}
\langle c^\dagger_{n+r}(t)c_n(0)\rangle_{L,N} & = & f_{L,N}(r,t) \\
\label{basiccorr2}
\langle c_{n+r}(t)c^\dagger_n(0)\rangle_{L,N} & = & (-1)^{r} f_{L,L-N}(-r,t) 
\eea
and therefore
\be 
S_\rho(r,t) = (-1)^{r} f_{L,N}(r,t) f_{L,L-N}(-r,t).
\ee 

Now we study the thermodynamic limit $L,N\to\infty$ such that $\rho=N/L$
is finite. We recall the Fermi momentum $k_F = \pi \rho$ and introduce the 
continuum dispersion relation
$\epsilon(p) := - \rme^{-ip}$
with its real and imaginary parts 
$\epsilon_1(p) := \Re(\epsilon(p)) = - \cos{p}, 
\epsilon_2(p) := \Im(\epsilon(p)) =  \sin{p}$.
By taking the derivative of the continuum dispersion relation
w.r.t. the momentum $p$ one obtains the Fermi velocity
$v_F = \epsilon_1'(k_F) = \sin{k_F} = \sin{(\pi\rho)}$
and the collective velocity
$v_c = \epsilon_2'(k_F) = \cos{k_F} = \cos{(\pi\rho)}$.
Notice that in the Hermitian case $\epsilon_2(p) = 0$ and therefore $v_c=0$.

We also define the complex coordinate
\bel{def:z}
z := i r+ \epsilon'(k_F) t = i \tilde{r} + v_F t  = v_F t (1+i \xi)
\ee
where $\tilde{r} = r - v_c t$ and the phase angle
\bel{def:phase}
\varphi(r,t)  := k_F r - \epsilon_2(k_F) t  = k_F r - \sin{(k_F)} t .
\ee
Scaling analysis of \eref{fdef} then yields
\be 
f_\rho(r,t) = \frac{\rme^{-i\varphi(r,t)- t \cos{(k_F)}}}{2\pi \bar{z}} + c.c.
\ee
and we arrive at
\bea 
\label{strucfunasympz}
S_\rho(r,t) & = & 
%\frac{1}{4\pi^2} \left(\frac{\rme^{-i\varphi(r,t)}}{\bar{z}} +
%\frac{\rme^{i\varphi(r,t)}}{z} \right)
% \left(\frac{\rme^{i\varphi(r,t)}}{\bar{z}} +
%\frac{\rme^{-i\varphi(r,t)}}{z} \right) \\
%\label{strucfunasympz}
%& = & 
\frac{1}{4\pi^2} \left( \frac{1}{z^2} + \frac{1}{\bar{z}^2}
+ \frac{2\cos{(2\varphi(r,t))}}{z\bar{z}} \right) \\
\label{strucfunasympxi}
& = &  \frac{1}{2(\pi v_F t)^2}\left[ \frac{1-\xi^2}{(1+\xi^2)^2} 
+ \frac{\cos{(2\varphi(r,t))}}{1+\xi^2} \right]
\eea
which is of the predicted form \eref{1} with $v_L=v_F$ and shows that in 
the Hermitian
case where $\epsilon_2(p)=0$ one has $v_c=\omega=0$ as in usual CFT.

Next we consider open boundary conditions. For maximal positive current the back-hopping rates proportional to $q$ are irrelevant and only
injection to site 1 with rate $\alpha  p$ 
and absorption into a reservoir at site $L$ 
with rate $\beta  p$ need to be considered.  Nevertheless,
open boundaries create several technical difficulties for the exact treatment, viz.
lack of periodicity, violation of particle conservation and loss of the 
bilinear free-fermion property underlying the computations of the previous 
paragraphs. The latter problem, however, can be overcome by 
augmenting the lattice with auxiliary boundary sites 
$0$ and $L+1$ which swap their state (empty or occupied) whenever a 
creation or annihilation event occurs. This leaves the dynamics of the 
exclusion process unchanged. 

The dependence on the boundary rates $\alpha$ and $\beta$ can be
removed by the similarity transformation $V = \prod_{k=1}^L u_k^{\hat{n}_k}$.
with the choice $p = (2\alpha\beta)^{-\frac{1}{L+1}}$ and
$u_k = \sqrt{2} \alpha p^k$. The transformed generator then reads
in terms of Jordan-Wigner fermions
\bel{genopenff}
H = - \sum_{k=1}^{L-1} c_{k+1}^\dagger c_k 
- \frac{1}{\sqrt{2}}\left[c_1^\dagger(c_0 -c_0^\dagger) + (c_{L+1}+c_{L+1}^\dagger)c_L\right].
\ee
For $L$ even the lack of periodicity and particle number conservation is
overcome by the Bogolyubov transformation 
\bea
\label{Bogogen}
b_k & = & \frac{1}{\sqrt{2}} \left(c_k+ (-1)^k c_{L+1-k}^\dagger\right), \quad
1\leq k \leq L \\
b_0 & = & \frac{1}{2} \left(c_0-c_0^\dagger + c_{L+1}+c_{L+1}^\dagger \right) .
\eea
The equations of motion read $\comm{H}{b_k} = b_{k-1}$ with periodic boundary
conditions, even though the original problem has no translation invariance.
Subsequent Fourier transformation
%\be 
%\hat{b}_p = \frac{1}{\sqrt{L+1}} \sum_{n=0}^L \rme^{\frac{-2\pi i pn}{L+1}} b_n
%\ee
%which also satisfy fermionic anticommutation relations 
leads to
$H = \sum_{p=0}^L \epsilon_p \hat{b}_p^\dagger \hat{b}_p$
with the single-particle ``energies''
$\epsilon_p = - \rme^{-\frac{2\pi i p}{L+1}}$.

Thus we are back to the periodic problem, albeit with $L+1$ sites
and only odd particle number which follows from the boundary conditions
in \eref{genopenff}.
The ground state is given by populating all negative energy modes 
$\Re(\epsilon_p) = − \cos{(2\pi p/(L+1))} \leq 0$ and therefore 
\bel{eps0open}
- e_0^{open} = \frac{1}{\pi} + \frac{1}{\pi L} + \frac{\pi}{24 L^2} + O(L^{-3}).
\ee
For the lowest energy gap 
one finds the leading finite-size correction for the real part
\bel{gaprealopen}
\Re{(\delta^{open}}) = \frac{4\pi}{L} + O(L^{-2})
\ee
which is twice the value of the periodic system at half-filling
where $v_F=1$.
%Both results are in agreement with boundary CFT with $c=1$ and
%boundary critical exponent $x=2$ \cite{Affl86,Bloe86}.

The stationary density of the conditioned process is constant with $\rho_n=1/2$
as one would expect for maximal current or activity. For the dynamical structure
function one can repeat the analysis of the periodic system due to the pseudo-
periodicity of \eref{Bogogen}. However, we need the full Wick theorem since
the second part is non-zero because of the lack of particle number
conservation. After some computation using the Bogolyubov transformation 
\eref{Bogogen} and taking the thermodynamic limit one arrives at
%\bea
%\langle c^+_n(t) c_m\rangle &=& \frac{1}{2}\left[ f_{L+1,2{N^\ast}+1}(n-m,t) + f_{L+1,L-2{N^\ast}}(n-m,t)\right]\\
%\langle c^+_n(t) c^+_m\rangle & = & \frac{(-1)^{m+1}}{2} \left[ f_{L+1,2{N^\ast}+1}(n+m,t) +  f_{L+1,L-2{N^\ast}}(n+m,t)  \right].
%\eea
%S_L(n,m,t) = S_{L+1,\lceil L/2\rceil }(n-m,t) - S_{L+1,\lceil L/2\rceil}(n+m,t)
%\ee 
%which in the thermodynamic limit yields
\be 
\label{dynstrucfunopen1}
S^{open}(n,m,t) = S_{1/2}(n-m,t) - S_{1/2}(n+m,t).
\ee 

Intriguingly the ground state results \eref{eps0} and \eref{gapreal} for the
periodic system and \eref{eps0open} and \eref{gaprealopen} for the open
system can be understood in terms of CFT 
even though \eref{H1} and \eref{genopenff} are strongly non-Hermitian. 
After rescaling time by the Fermi velocity $v_F$
(identified as such by the computation of the dynamical structure function)
and using finite-size scaling theory for conformal invariance \cite{Affl86,Bloe86}
the leading corrections \eref{eps0} and \eref{eps0open} to the
ground state energy correspond to central charge $c=1$ of the Virasoro algebra.
The real part $2\pi x/L$ of the lowest energy gap corresponds to the lowest 
critical bulk exponent $x=1$ of a primary field of the corresponding CFT for
the periodic case and $x=2$ for the open boundary conditions,
in complete analogy to the CFT describing the Hermitian $XX$-chain at half-filling
$\rho=1/2$. The non-hermitian nature of the time evolution enters only through 
the imaginary part of the energy gap which yields the collective velocity
$v_c = \cos{(\pi\rho)}$ and the oscillation frequency $\omega = \sin{(\pi\rho)}$.

The same CFT describes other quantum spin chains such as the spin-$s$ Heisenberg 
chain in the gapless regime \cite{Alca92a,Alca97} with anisotropy
$0\leq \Delta<1$ whose non-hermitian generalization maps to the 
partial exclusion process \cite{Schu94} conditioned an atypical current and
activity. One has an energy gap for integer spin (Haldane conjecture)
only for negative $\Delta$. In fact, it opens up exactly at $\Delta=0$ 
where the mapping to stochastic dynamics gets lost, as
positivity of the transition rates implies $\Delta \geq 0$.
The energy gap is finite in the ferromagnetic regime $\Delta > 1$, 
corresponding to phase separation in the particle system. These observations
support the notion of universality of \eref{1} and also confirm the 
dynamical phase transition to a phase-separated regime at low current or activity.

The dynamical structure function can be understood in terms of CFT
by adapting the standard splitting of the free-fermion operators $c_n(t)$ 
into right-movers (with positive momentum) and left-movers 
(with negative momentum) in the hermitian case \cite{Giam04} to the 
non-hermitian scenario. With
$\varphi(n,t) = k_F n - \Im(\epsilon_{k_F})t $ and 
$\varepsilon = \Re(\epsilon_{k_F})$
this yields
\be 
c_n(t) = \frac{1}{\sqrt{2\pi}} \left[\rme^{-i \varphi(n,t) + 
\varepsilon t} \psi_{\tilde{n}}(t)
+ \rme^{i \varphi(n,t) - 
\varepsilon t} \bar{\psi}_{\tilde{n}}(t)\right]
\ee
and similar for $c^\dagger_n(t)$ with Galilei-transformed coordinate
$\tilde{n}=n-v_ct$. For large $n$ and $t$ the two-point 
functions are predicted from CFT  with central charge $c=1$ to be 
\bel{confcorr}
\langle \psi^\dagger(z) \psi(z') \rangle = \frac{1}{z-z'}, \quad
\langle \bar{\psi}^\dagger(\bar{z}) \bar{\psi}(\bar{z}') \rangle = \frac{1}{\bar{z}-\bar{z}'}.
\ee
This then leads to the dynamical structure function with static
exponent $\gamma=1$ that we computed.
The same scaling form \eref{1} with a non-universal exponent 
$\gamma > 1$ can be obtained via bosonization \cite{Giam04}
for non-maximal speeding-up, i.e., $\Delta > 0$. Since $t$ is
to be understood in units of the lattice spacing the oscillating
part thus becomes irrelevant even without spatial course-graining.
The dynamical structure function for open boundaries can be understood 
from boundary conformal field theory with reflecting boundary conditions
which yields non-vanishing correlators 
between the right- and left-movers which are of the form similar to
\eref{confcorr}, but with a $1/(z+z')$ dependence coming from reflection.

From \eref{XXZhopping} we read off the remarkable fact
that for $\Delta=\cosh{(f+\lambda)}$, i.e., at a critical activity parameter
\bel{muc}
\mu_c = - \ln{\left(\frac{\cosh{(f+\lambda)}}{\cosh{f}}\right)}
\ee
the conditioned generator becomes the generator of an unconditioned 
ASEP with hopping asymmetry $f' = f+\lambda$. Then for $\mu < \mu_c$ 
one has phase separated 
atypical behaviour as shown in \cite{Leco12} for $f=\lambda=0$ and
in \cite{Spoh99} for $f=0$,
whereas for $\mu > \mu_c$  one has -- as elaborated above --
conformal invariance. On the phase transition line $\mu=\mu_c$ 
one has typical stochastic dynamics with $z=2$ for $\lambda=-f$ (symmetric
simple exclusion process) and
$z=3/2$ for $\lambda \neq -f$ (ASEP).

\begin{figure}[ptb]
\begin{center}
\includegraphics[scale=0.3]{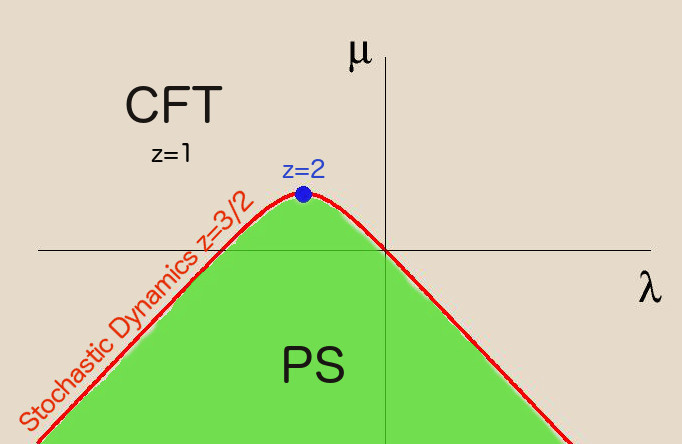}
\end{center}
\caption{(Color online)
Phase diagram of the conditioned ASEP as a function of
current parameter $\lambda$ and the activity parameter $\mu$
for positive drive $f>0$. The exact phase transition line that
separates the conformally invariant regime CFT from the phase-separated
regime PS is given by \eref{muc}.}%
\label{Fig:ASEP2}%
\end{figure}

Generally we argue that the phase diagram of atypical behaviour in 
$d$-dimensional driven diffusive systems far from thermal equilibrium
can be explored by studying the critical behaviour of $d+1$-dimensional equilibrium
models. In $1+1$ dimension with a
single conserved species there is a dynamical phase transition from a 
dynamical critical regime for fast dynamics (described by conformal
field theory) through a phase transition line corresponding to typical dynamics
(described by non-linear fluctuating hydrodynamics) 
to a phase-separated regime of slow dynamics (accessible by macroscopic
fluctuation theory). This demonstrates that large fluctuations in driven
diffusive systems cannot be understood by some simple upscaling procedure but is 
sustained by spatio-temporal correlations that are 
qualitatively different from typical behaviour.

For systems with more than one conservation law the 
scenario is more complex. It has been shown
recently that typical behaviour of stochastic particle 
systems with local dynamics have dynamical critical exponents $z_i>1$ given by 
the ratios of neighboring Fibonacci numbers $F_{i+1}/F_{i}$, starting with 
$F_2=1$ and $F_3=2$ \cite{Popk15}. Also, quantum systems
for $n$ conserved species that would describe the critical behaviour in the 
regime of high current or activity may be governed by a CFT with 
central charge $c>1$. For low current or activity one still expects a 
phase-separated regime, which, however, may exhibit richer behaviour than 
for $n=1$.

G.M.S. thanks the University of Lorraine (Nancy) and the University of S\~ao Paulo
for kind hospitality. F.C. Alcaraz and J. Dubail are gratefully acknowledged 
for useful discussions. This work was 
supported by Deutsche Forschungsgemeinschaft.


\begin{thebibliography}{99}

\bibitem{Spoh91} 
H. Spohn,
{\it Large Scale Dynamics of Interacting particles} 
(Berlin, Springer, 1991).

\bibitem{Schu01} 
G.M. Sch\"utz,
%Exactly solvable models for many-body systems far from equilibrium, 
in: Domb C. and Lebowitz J. (eds.), Phase Transitions and
Critical Phenomena Vol. 19, 1--251. (Academic Press, London, 2001).

\bibitem{Scha10}
A. Schadschneider, D. Chowdhury, and K. Nishinari,
%Schadschneider A Chowdhury D and Nishinari K 2010
{\it Stochastic Transport in Complex Systems}
(Elsevier, Amsterdam, 2010).

\bibitem{Derr07}
B. Derrida,  
%Non-equilibrium steady states: fluctuations, large deviations of the density, 
%of the current  
J. Stat. Mech. P07023 (2007).

\bibitem{Spoh14}
H. Spohn,
%Nonlinear Fluctuating hydrodynamics for anharmonic chains.
J. Stat. Phys. {\bf 154}, 1191--1227 (2014).

\bibitem{Card96}
J.L. Cardy, 
\textit{Scaling and Renormalization in Statistical Physics} 
(Cambridge University Press, Cambridge, 1996).

\bibitem{Henk99}
M. Henkel, 
\textit{Conformal Invariance and Critical Phenomena}
(Springer, Berlin, 1999).

\bibitem{Bert15}
L. Bertini, A. De Sole, D. Gabrielli, G. Jona Lasinio, and C. Landim,
%Bertini, L., De Sole, A., Gabrielli, D., Jona Lasinio, G., Landim, C.:
% Macroscopic fluctuation theory.
Rev. Mod. Phys. \textbf{87}, 593--636 (2015).

\bibitem{Bodi05}
T. Bodineau and B. Derrida, 
%Distribution of current in nonequilibrium diffusive systems and phase transitions. 
Phys. Rev. E \textbf{72}(6), 66110 (2005).

\bibitem{Espi13}
Espigares, C.P., Garrido, P.L., Hurtado, P.I.:
%Dynamical phase transition for current statistics in a simple driven diffusive system.
Phys. Rev. E \textbf{87}, 032115 (2013).

\bibitem{Bodi06}
T. Bodineau and B. Derrida, 
%Current Large Deviations for Asymmetric Exclusion Processes with Open Boundaries, 
J. Stat. Phys.  \textbf{123}(2), 277--300 (2006).

\bibitem{Beli13b} Belitsky, V., Sch\"utz, G.M.: 
%Antishocks in the ASEP with open boundaries conditioned on low current.
J. Phys. A: Math. Theor. \textbf{46}, 295004 (2013).

\bibitem{Laza15}
A. Lazarescu,
%The Physicist's Companion to Current Fluctuations: One-Dimensional Bulk-Driven Lattice Gases,
arXiv:1507.04179 (2015).

\bibitem{Leco12}
Lecomte, V., Garrahan, J.P., Van Wijland, F.:
%Inactive dynamical phase of a symmetric exclusion process on a ring.
J. Phys. A: Math. Theor. \textbf{45}, 175001 (2012).

\bibitem{Schu15a}
G.M. Sch\"utz,
%The Space-Time Structure of Extreme Current and Activity Events in the ASEP,
in: Springer Proceedings of the 
``International School and Workshop on Nonlinear Mathematical Physics and Natural Hazards'',
B. Aneva and M. Kouteva-Guentcheva (eds.), (Springer, Cham, 2015).

\bibitem{Chet14}
Chetrite, R., Touchette, H.:
%Nonequilibrium Markov processes conditioned on large deviations. 
Ann. Henri Poincar{\'e}, \textbf{16}(9), 2005--2057 (2015)

\bibitem{Jack15}
R. L. Jack and P. Sollich, 
%Effective interactions and large deviations in stochastic processes,
Eur. Phys. J. Spec. Top. \textbf{224}, 2351--2367 (2015).

\bibitem{Baxt82}
Baxter, R.J.:
Exactly Solved Models in Statistical Mechanics. 
Academic, New York (1982)

\bibitem{Popk10} 
Popkov, V., Simon, D., Sch\"utz, G.M.:
%ASEP on a ring conditioned on enhanced flux.
J. Stat. Mech. P10007 (2010).

\bibitem{Popk11} 
Popkov, V., Sch\"utz, G.M.:
%Transition probabilities and dynamic structure factor in the 
%ASEP conditioned on strong flux.
J. Stat. Phys.  \textbf{142}(3), 627--639 (2011).

\bibitem{Spoh99} 
H. Spohn,
%Bosonization, vicinal surfaces, and hydrodynamic fluctuation theory.
Phys. Rev. E \textbf{60}, 6411-6420 (1999).

\bibitem{Lieb61}
Lieb, E., Schultz, T., Mattis, D.: 
%Two Soluble Models of an Antiferromagnetic Chain.
Ann. Phys. 16, 407--466 (1961)

\bibitem{Affl86}
Affleck I,
%Universal term in the free energy at a critical point and the conformal anomaly,
Phys. Rev. Lett. \textbf{56}, 746--749 (1986)

\bibitem{Bloe86}
Bl\"ote H W J, Cardy J L and Nightingale M P, 
%Conformal invariance, the central charge, and universal finite-size amplitudes at criticality,
Phys. Rev. Lett. \textbf{56}, 742--745 (1986)

\bibitem{Alca92a}
F. C. Alcaraz and A. Moreo,
%Critical behavior of anisotropic spin-S Heisenberg chains.
Phys. Rev. B \textbf{46}, 2896 (1992)

\bibitem{Alca97}
F. C. Alcaraz, A. L. Malvezzi 
%Critical Behaviour of Mixed Heisenberg Chains
J. Phys. A: Math. Gen. \textbf{30}(3) 767 (1997)

\bibitem{Schu94}
G. Sch\"utz, S. Sandow,
%Sch\"utz, G., and Sandow, S.:
%Non-abelian symmetries of stochastic processes: derivation of correlation functions 
%for random vertex models and disordered interacting many-particle systems.
Phys. Rev. E \textbf{49}, 2726--2744 (1994).

\bibitem{Giam04}
T. Giamarchi,
\textit{Quantum Physic in One Dimension}
(Oxford University Press, Oxford, 2004).

\bibitem{Popk15} 
V. Popkov, A. Schadschneider, J. Schmidt, G.M. Sch\"utz,
%Fibonacci family of dynamical universality classes,
Proc. Natl. Acad. Sci. USA \textbf{112}(41) 12645-12650 (2015).



%%%%%%%%%%%%%%%


\end{thebibliography}
\end{document}